\documentclass{article}
\usepackage{stmaryrd}
\usepackage{amsmath}
\usepackage{amssymb}
\usepackage{listings}
\usepackage{ulem}
\renewcommand{\emph}[1]{\textit{#1}}
\usepackage{fullpage}
\usepackage{verbatim}
\usepackage{wasysym}
\usepackage{booktabs}
\usepackage{xspace}
\usepackage{authblk}

\let\olditemize=\itemize
\def\itemize{
	\olditemize
	\setlength{\itemsep}{-1ex}
}
\let\oldenumerate=\enumerate
\def\enumerate{
	\oldenumerate
	\setlength{\itemsep}{-1ex}
}

\newcommand{\ecos}{eCos\xspace}

\newcommand{\Id}{\textsf{Id}}
\newcommand{\Flavors}{\textsf{Flavors}\xspace}
\newcommand{\Kinds}{\textsf{Kinds}\xspace}
\newcommand{\FlBool}{\textsf{bool}\xspace}
\newcommand{\FlNone}{\textsf{none}\xspace}
\newcommand{\FlBoolData}{\textsf{booldata}\xspace}
\newcommand{\FlData}{\textsf{data}\xspace}
\newcommand{\id}{\textsf{id}}
\newcommand{\Cdl}{\textsf{Cdl}}

\newcommand{\Nodes}{\textsf{Nodes}}
\newcommand{\Exp}{\text{\textit{Exp}}}
\newcommand{\LExp}{\text{\textit{LExp}}}
\newcommand{\Powerset}[1]{\ensuremath{\mathcal{P}(#1)}}
\newcommand{\Confs}{\textsf{Confs}}
\newcommand{\Data}{\textsf{Data}}
\newcommand{\const}{\textsf{const}}
\newcommand{\access}{\textit{access}}
\newcommand{\bool}{\textit{bool}}

\newcommand{\eval}{\textit{eval}}

\newcommand{\CTC}{\textsf{CTC}}

\newcommand{\denotation}[1]{\ensuremath{[\![#1]\!]}}
\newcommand{\cdlDenotation}[1]{\denotation{#1}_{\textsf{cdl}}}
\newcommand{\nodeDenotation}[1]{\denotation{#1}_{\textsf{node}}}
\newcommand{\flavorDenotation}[1]{\denotation{#1}_{\textsf{flavor}}}
\newcommand{\calculatedDenotation}[1]{\denotation{#1}_{\textsf{calculated}}}
\newcommand{\lvDenotation}[1]{\denotation{#1}_{\textsf{legal\_values}}}
\newcommand{\interfaceDenotation}[1]{\denotation{#1}_{\textsf{interface}}}

\newcommand{\booleanOps}{\ensuremath{\mid\mid,\&\&,\textit{implies},\textit{eqv},\textit{xor}}}
\newcommand{\arithmOps}{\ensuremath{+,-,*,/,\%,<<,>>,\text{\^{}},\&,\mid}}
\newcommand{\otherOps}{\ensuremath{==,\text{!=},<,>,<=,>=}}
\newcommand{\Func}{\textsf{Func}}
\newcommand{\functions}{\textit{get\_data, is\_active, is\_enabled, is\_loaded, is\_substr, is\_xsubstr, version\_cmp}}

\newcommand{\calculated}{\texttt{calculated}}

\newcommand{\pCdlDenotation}[1]{\denotation{#1}_{\textsf{pCdl}}}
\newcommand{\pNodeDenotation}[1]{\denotation{#1}_{\textsf{pNode}}}
\newcommand{\pFlavorDenotation}[1]{\denotation{#1}_{\textsf{pFlavor}}}
\newcommand{\pCalculatedDenotation}[1]{\denotation{#1}_{\textsf{pCalculated}}}
\newcommand{\pInterfaceDenotation}[1]{\denotation{#1}_{\textsf{pInterface}}}
\newcommand{\peval}{\textit{$eval_p$}}
\DeclareMathOperator*{\bigxor}{\mathbf{XOR}}
\newcommand{\BExp}{\textit{BExp}}
\newcommand{\rewrite}{\textit{rewrite}\xspace}

\newcommand{\pBooleanOps}{\ensuremath{\mid\mid,\&\&,\textit{implies},\textit{eqv}}}

\title{Formal Semantics of the CDL Language \\[.2\baselineskip]\normalsize\textbf{Technical Note}}
\author[1]{Thorsten Berger}
\author[2]{Steven She}
\affil[1]{\texttt{berger@informatik.uni-leipzig.de}, University of Leipzig, Germany}
\affil[2]{\texttt{shshe@uwaterloo.ca}, University of Waterloo, Canada}

\date{January 2010}

\begin{document}
\maketitle
%\label{EcosCDLSemantics}
\begin{abstract}We reverse-engineer a formal semantics of the \textit{Component Definition Language} (CDL), which is part of the highly configurable, embedded operating system \ecos\footnote{http://ecos.sourceware.org}. This work provides the basis for an analysis and comparison of the two variability-modeling languages Kconfig and CDL\,\cite{berger.ea:2012:tr,berger2010variability}. The semantics given in this document are based on analyzing the CDL documentation,\footnote{http://ecos.sourceware.org/docs-3.0/cdl-guide/cdl-guide.html} inspecting the source code of the toolchain as well as testing the tools on particular examples.
\end{abstract}

\section{Semantics}

\subsection{Abstract Syntax}
\paragraph{Features, types and constraints.}
Let \Id\ be a finite set of features, let \Kinds be a set of domain-specific feature kinds and let \Flavors\ be a set of types that further define a feature's possible values. More precisely, $\Kinds=\{$\textsf{package, component, option, interface}$\}$ and $\Flavors=\{\FlNone, \FlBool, \FlBoolData, \FlData\}$. Furthermore, we introduce two types of expressions allowed in CDL: \textit{Goal expressions} and \textit{list expressions}.

Concerning the first one, we define $\Exp(\Id)$ to be a set of goal
expressions over \Id, generated by the following grammar:
\begin{equation}
	e ::= \id
	\mid \const
	\mid e \otimes e 
	\mid !e
	\mid \text\textasciitilde e
	\mid e \oplus e
	\mid e \oslash e
	\mid \Func(e,e,...)
	\mid e ? e : e
\end{equation}

Here, $\otimes\in\{\booleanOps\}$, $\oplus\in\{\arithmOps\}$, $\oslash\in\{\otherOps\}$, $\Func\in$\{\functions\}\label{cdlfunctions}, $\id\in\Id$ and $\const\in\Data$, whereas \Data\ is a set of untyped data (say all character strings).

The second type of expressions, so-called list expressions represent an enumeration of values or ranges, which can be computed by goal expressions. Thus, we define $\LExp(\Id)$ to be a set of list
expressions over goal expressions, generated by the following grammar ($e\in\Exp(\Id)$):
\begin{equation}
	l ::= \left( e \mid e \textsf{ to } e \right) [\ \text{\textvisiblespace} l\ ]
\end{equation}

\paragraph{CDL models.}
\Cdl\ is the set of all possible models
in CDL.  Each CDL model $m\in\Cdl$ is a
set of nodes, so $\Cdl = \Powerset\Nodes$, where
\begin{equation}
	\Nodes = \Id \times \lceil\Id\rceil \times \Flavors \times \Powerset{\Exp(\Id)} \times \Powerset{\Exp(\Id)} \times \lfloor\Exp(\Id)\rfloor \times \lfloor\LExp(\Id)\rfloor \times\Kinds \times \Powerset{\Id}
\end{equation}

If $(n,p,fl,ai,req,cl,lv,knd,imp) \in \Nodes$, then $n$ is the name, $p$ is the parent of the node ($p=\top$ for nodes at the top level), $ai$
is a set of \texttt{active\_if} visibility goal expressions, and $req$ is a set of
\texttt{requires} goal expressions. Further, $cl$ denotes a \calculated\ goal expression that prescribes the feature's values and $lv$ is a \texttt{legal\_values} list expression restricting its values. Finally, $knd$ specifies the node's domain-specific kind and $imp$ specifies whether the node implements one or more interfaces. There is no further restriction on both values, that is, an interface can even implement other interfaces.
We write $\Id(m)$ to denote names of nodes in the model $m$, so $\Id(m) = \{
n \mid (n,\_,\_,\_,\_,\_,\_,\_,\_)\in m \}$.

\paragraph{Well-formedness.}
\label{par:wellformedness}
CDL introduces some more constraints on the syntax of the model. If

$(n,p,fl,ai,req,cl,lv,knd,imp) \in \Nodes$, it has to fulfill the following invariants:
\begin{itemize}
	\item $fl=none \rightarrow cl=\bot$ (\texttt{calculated} has no effect if flavor is \FlNone)
	\item $cl\neq\bot \rightarrow lv=\bot$ (\texttt{calculated} and \texttt{legal\_values} exclude each other)
	\item $fl \in \{bool\} \rightarrow lv=\bot$ (\texttt{legal\_values} applies to nodes with non-fixed data value only, see Eq.\ref{eq:domain})
	\item $knd=\textit{interface} \rightarrow ( fl\neq\FlNone \wedge cl=\bot )$ (Interfaces must neither have the \FlNone\ flavor nor a \texttt{calculated} property)
	%\item Packages always have flavor \FlBoolData\, are enabled and have their version as data value.
	\item The parent relationship $p$ should define a tree, with the virtual $\top$ as the root. Furthermore, nodes of kind \textit{option} must not be parents of other nodes.
\end{itemize}

\paragraph{Pre-processing notes.}
For convenience and conciseness reasons, the abstract syntax given in this section depends on the following preprocessing steps from the concrete syntax:
\begin{enumerate}
	\item Similar to the configuration tool, we introduce a synthetic root element $\top$, which is a parent to every top-level package.
	\item In case no flavor is specified for a node, we set the flavor ($fl$) property (according to the CDL documentation) to \FlBoolData for packages, to \FlBool for components and options, and to \FlData for interfaces.
	\item The \texttt{requires}, \texttt{active\_if} and \texttt{calculated} properties can contain an enumeration of goal expressions separated by whitespace. We convert such enumerations to a disjunction of their goal expressions.
\end{enumerate}

\subsection{Semantic Domain}

A configuration is an assignment of triples of values to nodes.  The set of
all possible configurations is:
\begin{equation}
\label{eq:domain}
	\Confs = \lceil\Id\rceil \rightarrow (\{0,1\} \times \{0,1\} \times \Data)
\end{equation}

If $c\in\Confs$ and $x\in\Id$, we write $c(x)_1$ for the first component of the
valuation (the \emph{enabled state}), $c(x)_2$ for the second one (the \emph{enabled value}), and $c(x)_3$ for the third component of the valuation (the \emph{data value}). The first component specifies whether the node is actually in the configuration, that is, whether it influences the build of \ecos in some sense. The latter two components refer to values the user can give to a node. We predefine the valuation of the $\top$ element as follows: $c(\top)_1=1, c(\top)_2=1, c(\top)_3=1$.

Now, a semantics of a CDL model is given in
terms of sets of configurations, so \Powerset{\Confs} is our semantic domain and the semantic function has the signature:
\begin{equation}
	\cdlDenotation\cdot : \Cdl \rightarrow \Powerset\Confs\\
\end{equation}

\begin{equation}
	\denotation\cdot: \textsf{Model} \rightarrow \Powerset\Confs\\
\end{equation}

\subsection{Semantics}

\paragraph{Helper functions.}
Let $\access : \Id \times \Confs \rightarrow \Data$ denote a function that returns the value of a feature under a certain configuration while taking its enabled state into account.
\begin{equation} access(x,c) =
 \begin{cases}
  0   &  \text{iff~}c(x)_1=0\\
  c(x)_3 & \text{iff~}c(x)_1=1
\end{cases}
\end{equation}

Since arbitrary values can be returned for a feature's occurrence in an expression, and since they can be direct input to Boolean operators (e.g., ``feature A requires B \&\& C'' and C could have flavor data or booldata), we define a cast of arbitrary values to Boolean values in the TCL/TK style. More precisely, $\bool: \Data \rightarrow \{0,1\}$. Please note that $\bool$ is also defined for plain Boolean values ($\{0,1\}\subset\Data$), which are the return type if nodes are inactive, disabled, or bool.
\begin{equation} \bool( v ) = 
	\begin{cases}
		0 & \text{iff~} v=\text{0+} \vee v=\text{""+} \\
		1 & \text{otherwise}
	\end{cases}
\end{equation}

%Now, we continue with a satisfaction relation for expressions under a configuration $ \models : \Exp(\Id)\ \times \Confs \rightarrow \{0,1\}$, which is defined as follows:
%\begin{equation}
%c \models e \text{ iff } \bool(\eval(e,c)) = 1
%\end{equation}
For the evaluation of goal expressions, we define the function $\eval:\Exp(\Id)\times\Confs\rightarrow \Data$ recursively as follows, with $x\in \Id$, $e_1,e_2,e_3\in\Exp(\Id)$ and $\const\in \Data$:
\begin{align}
\begin{split}
\eval(x,c)= &\access(x,c) \\
\eval(\const,c)=&\const \\
\eval(!e_1,c)=&\textsf{non~}\bool(\eval(e_1,c)) \\
\eval(e_1 \otimes e_2,c)=& \phi_0(\bool(\eval(e_1,c)),\bool(\eval(e_2,c))) \\
&\text{ with } \phi_0=\textsf{vel,et,seq,eq,aut} \text{ for }\otimes=\booleanOps \\
\eval(e_1 \oplus e_2,c)=& \phi_1(\eval(e_1,c),\eval(e_2,c)) \\
&\text{ with }\phi_1\text{ TCL's arithmetic for }\oplus=\arithmOps \\
\eval(e_1 \oslash e_2,c)=& \phi_2(\eval(e_1,c),\eval(e_2,c)) \\
&\text{ with }\phi_2\text{ TCL's comparison operators for }\oslash=\otherOps \\
\eval(e_1?e_2:e_3,c)=& \begin{cases}
\eval(e_2,c) & \text{iff~} \bool(\eval(e_1,c))\\
\eval(e_3,c) & \text{otherwise}
\end{cases}
\end{split}
\end{align}
We left out CDL's built-in functions (see \Func\ in \ref{cdlfunctions}) in the definition of \eval\ and refer to the CDL online documentation instead.

For the evaluation of values against the \texttt{legal\_values} property, we introduce a satisfaction relation $ \models : \Data\times\Confs\times\LExp(\Id) \rightarrow \{0,1\}$. For $d\in\Data; c\in\Confs; e_1,e_2\in\Exp(\Id), l_1,l_2\in\LExp(\Id)$, we define the relation:
\begin{align}
\begin{split}
d &\models_c e_1 \hspace{1.5cm} \text{ iff } d = \eval(e_1,c) \\
d &\models_c e_1 \textsf{ to } e_2 \hspace{0.62cm} \text{ iff } \bool(\eval( e_1 <= d\ \&\&\ d <= e_2, c )) \\
d &\models_c l_1\ \text{\textvisiblespace}\ l_2 \hspace{0.85cm} \text{ iff } (d,c) \models l_1 \vee (d,c) \models l_2 \\
\end{split}
\label{eq:listevaluation}
\end{align}

\paragraph{Semantic function.}
The semantics of a model is just an intersection of denotations of all the
nodes except that we need to ensure that all the unloaded packages (that is, their nodes) are
enforced to be false. Furthermore, we adhere to the semantics of the nodes' \texttt{flavor} as well as we take their \texttt{calculated}/\texttt{legal\_values} property and the interface concept into account.
\begin{align}
\begin{split}
\cdlDenotation m = &
\left(\bigcap_{n\in m} \nodeDenotation n \right)
\cap
\left(\bigcap_{n\in m} \flavorDenotation n \right)
\cap
\left(\bigcap_{n\in m} \calculatedDenotation n \right)
\cap
\left(\bigcap_{n\in m} \lvDenotation n \right) \\
& \cap
\left(\bigcap_{n\in m} \interfaceDenotation n \right)
\cap
\left\{~ c\! \in\! \Confs \mid c(x)_1=0 \text{ for all } x\in \Id\setminus\Id(m)
~\right\}
\end{split}
\end{align}

The semantics of a node is a set of all configurations that (1) contain the node's parent, (2) have the node's enabled value set, and (3) can satisfy the node's
constraints. Notably, we ignore the unloaded packages here, as we treated them
all globally above (here 'unloaded' = 'not mentioned in the model'). For brevity, we also introduce a macro for cross-tree constraints: $\CTC=\forall e\in ai\cup req.\bool(\eval(e,c))$.

\begin{equation}
	\nodeDenotation{(n,p,\_,ai,req,\_,_,\_,\_)} =  \{ c \in \Confs \mid c(n)_1 \leftrightarrow (c(p)_1 \wedge c(n)_2 \wedge \CTC ) \}
	\label{eq:nodeDenotation}
\end{equation}

Next, we continue with the denotation of nodes according to their flavor property. The values \FlNone and \FlData are the equivalent to mandatory features in Feature Modeling (FM) with one big difference: In CDL, such nodes can be made optional with cross-tree constraints, whereas in FM, cross-tree constraints of mandatory features also constrain the parent. In the flavor denotation, we set the \textit{enabled value} for \FlNone and \FlData features. With regard to Eq. \ref{eq:nodeDenotation}, such mandatory nodes still require the parent node present and cross-tree constraints satisfied.
\begin{align}
\begin{split}
\flavorDenotation{(n,p,(\FlNone|\FlData),ai,req,\_,\_,\_,\_)}&=\{c\in\Confs \mid c(n)_2\} \\
\flavorDenotation{(\_,\_,(\FlBool|\FlBoolData),\_,\_,\_,\_,\_,\_)}&=\Confs
\end{split}
\end{align}

The \calculated\ property forces a node's \emph{data} and \emph{enabled value} depending on the flavor. Please note that CDL excludes the \FlNone flavor for calculated nodes (cf. Sec. \ref{par:wellformedness}, well-formedness). For $cl\neq\bot$, we define:
\begin{align}
\begin{split}
\calculatedDenotation{(n,\_,\FlBool,\_,\_,cl,\_,\_,\_)}&=\{c\in\Confs\mid c(n)_2=\bool(\eval(cl,c))\} \\
\calculatedDenotation{(n,\_,\FlBoolData,\_,\_,cl,\_,\_,\_)}&=\{c\in\Confs\mid c(n)_3=\eval(cl,c) \wedge c(n)_2=\bool(c(n)_3) \} \\
\calculatedDenotation{(n,\_,\FlData,\_,\_,cl,\_,\_,\_)}&=\{c\in\Confs\mid c(n)_3=\eval(cl,c)\}
\end{split}
\end{align}

The \texttt{legal\_values} property restricts the data value of a node with a list expression. We define its denotation by using our satisfaction relation from Eq. \ref{eq:listevaluation}. Interestingly, the \texttt{legal\_values} property only excludes the flavor \FlBool by well-formedness rules (cf. Sec. \ref{par:wellformedness}), whereas it does not have any effect on \FlNone -flavored nodes. For $lv\neq\bot$, we define:
\begin{align}
\begin{split}
\lvDenotation{(n,\_,(\FlBoolData\mid\FlData),\_,\_,\_,lv,\_,\_)} &= \{c\in\Confs\mid c(n)_3 \models_c lv \}\\
\lvDenotation{(n,\_,\FlNone,\_,\_,\_,lv,\_,\_)} &= \Confs
\end{split}
\end{align}

Finally, we specify the denotation of interfaces, which represent derived features in CDL. However, please note that the \FlNone flavor is excluded by well-formedness rules (cf. Sec. \ref{par:wellformedness}).
\begin{align}
\begin{split}
\interfaceDenotation{(n,\_,\FlBoolData,\_,\_,\_,\_,\textit{interface},\_)} &= \{c\in\Confs\mid c(n)_3 = \mid impls(n,c)\mid \wedge \\
& \hspace{2.34cm} c(n)_2=\bool(c(n)_3)\}\\
\interfaceDenotation{(n,\_,\FlData,\_,\_,\_,\_,\textit{interface},\_)}&=\left\{c\in\Confs\mid c(n)_3 = \mid impls(n,c)\mid \right\} \\
\interfaceDenotation{(n,\_,\FlBool,\_,\_,\_,\_,\textit{interface},\_)}&=\left\{c\in\Confs\mid c(n)_2 = \bool(\mid impls(n,c)\mid) \right\}
\end{split}
\end{align}

where $impls:\Id\times\Confs\rightarrow\Powerset\Nodes$ is defined as follows:

\begin{equation}
impls(n,c)=\left\{x\in\Nodes \mid n\in x_{impl} \wedge c(x)_1=1 \right\}
\end{equation}

\section{Propositional Semantics}
We now describe a Boolean interpretation of the abstract syntax. Given CDL's expressiveness, there is no precise translation from a CDL model into propositional logic. While it is relatively easy to translate the hierarchy and flavor constraints into a propositional formula, this task becomes more complicated for cross-tree constraints and the interface concept. Our strategy is to approximate constraints as much as possible by loosening the original constraints, that is, the propositional semantics under-approximate the full ones.

We tailor the full semantics down to denotations that can be expressed in propositional logic. Based on these semantics, we implemented\footnote{https://code.google.com/p/variability/wiki/CDLTools} rewriting rules that take a full CDL model and convert it into a Boolean formula, enabling analysis based on SAT solvers. The latter comprises, for instance, satisfiability and dead feature checks, or building implication graphs.
\subsection{Propositional Semantic Domain}
We define
\begin{equation}
	\Confs_p = \lceil\Id\rceil \rightarrow \{0,1\}
\end{equation}
If $c_p\in\Confs_p$ and $x\in\Id$, we write $c_p(x)$ for the valuation of the node under a configuration. We also predefine $c_p(\top)=1$ for the $\top$ element. Similar to the full semantics, the propositional semantics of a CDL model is given in terms of sets of configurations, so \Powerset{\Confs_p} is our semantic domain. Thus, our semantic function has the following signature:
\begin{equation}
	\pCdlDenotation\cdot : \Cdl \rightarrow \Powerset{\Confs_p}\\
\end{equation}

Furthermore, we define some invariants between the full (\Confs) and propositional ($\Confs_p$) configuration spaces, basically answering the question: What does it mean if a feature under a Boolean configuration $c_p\in\Confs_p$ is true or false with regard to the full semantics $c\in\Confs$? Table \ref{tab:invariantsBetweenConfigurationSpaces} shows the invariants according to a node's flavor.

\begin{table*}[h]\centering
	\begin{tabular}{ll}
		Flavor & Invariant \\
		\midrule
		\FlBool & $c_p(n) = c(n)_1$ \\
		\FlNone & $c_p(n) = c(n)_1$ \\
		\FlBoolData & $c_p(n) = c(n)_1 \wedge c(n)_3\neq0$ \\
		\FlData & $c_p(n) = c(n)_1 \wedge c(n)_3\neq0$ \\
	\end{tabular}
	\caption{Invariants between configuration spaces}
	\label{tab:invariantsBetweenConfigurationSpaces}
\end{table*}

\subsection{Propositional Semantics}
\paragraph{Helper functions.}
A function such as $\access_p : \Id \times \Confs_p \rightarrow \{0,1\}$ is not necessary anymore, since a node does only have one value left (that is, $access_p(\id,c) = c_p(\id)$). However, a slightly different function $impls':\Id\times\Cdl\rightarrow\Powerset\Id$ will be helpful later in this section:
\begin{equation}
impls'(i,m) = \{n\in m \mid i \in n_{impl}\}
\end{equation}
\paragraph{Boolean expressions.}
We just have to consider goal expressions since list expressions only appear in \textit{legal\_values} constraints, which cannot be approximated\footnote{One could argue that it is possible to approximate special cases, such as \texttt{legal\_values} 0 and so on. However, it would spoil our translation with too many exceptions and we have not seen comparable examples in the real models.} in the propositional semantics since our semantic domain contains no \textit{data value} any more.
Let $\BExp(\Id)\subset\Exp(\Id)$ be a subset of Boolean expressions over \Id, which is defined by the following grammar, with $\otimes=\{\pBooleanOps\}$ and $\const\in\{0,1\}$:
\begin{equation}
	e ::= \id
	\mid \const
	\mid e \otimes e 
	\mid !id
\end{equation}

\paragraph{Boolean expression evaluation.} The evaluation of \BExp(\Id) now follows ordinary propositional semantics. Thus, the definition of a function $\peval:\BExp(\Id)\times\Confs_p\rightarrow\{0,1\}$ is pretty straightforward, with $x\in \Id$, $e_1,e_2\in\BExp(\Id)$ and $\const\in\{0,1\}$:
\begin{align}
\begin{split}
\peval(x,c_p)= &c_p(x) \\
\peval(\const,c_p)=&\const \\
\peval(!e_1,c_p)=&\textsf{non~}\peval(e_1,c_p) \\
\peval(e_1 \otimes e_2,c_p)=& \phi_0(\peval(e_1,c_p)),\peval(e_2,c_p)) \\
&\text{ with } \phi_0=\textsf{vel,et,seq,eq} \text{ for }\otimes=\pBooleanOps \\
\end{split}
\end{align}

\paragraph{Expression rewriting rules.}
Next, we define a partial function \rewrite:$\Exp(\Id)\times\Cdl\leadsto\BExp(\Id)$, which translates goal expressions from the full semantics to reduced Boolean ones. For $x\in\Id$, $m\in\Cdl$ and $e_1,e_2,e_3\in \Exp(\Id)$:
\begin{align}
\begin{split}
rewrite(x,m) &= \begin{cases}
x &\text{iff~} x\in\Id(m)\\
0 &\text{otherwise}\\
\end{cases}\\
rewrite( !x, m ) &= \neg rewrite(x,m)\\
rewrite(\const,m) &= \bool( \const )\\
rewrite(x=\const,m) &= \begin{cases}
rewrite(x,m) &\text{iff~} \bool(\const)\neq0\\
rewrite(\neg x,m) &\text{otherwise}\\
\end{cases}\\
rewrite(x>\const,m) &= \begin{cases}
rewrite(x,m) &\text{iff~} \const\in INT \wedge \const\geq 0\\
1 &\text{otherwise} \text{  (drop it)}\\
\end{cases}\\
rewrite(\textit{is\_substr}(x,\const),m) &= rewrite(x,m)\\
rewrite(e_1 \otimes e_2,m) &= rewrite(e_1,m) \otimes rewrite(e_2,m)\\
rewrite(e_1 ? e_2 : e_3,m) &= (rewrite(e_1,m) \rightarrow rewrite(e_2,m))\\
& \hspace{0.5cm} \wedge (\neg rewrite(e_1,m) \rightarrow rewrite(e_3,m))
\end{split}
\end{align}

For the interpretation of interfaces, we need to define a helper function $choose:\Powerset\Id\times\mathbb{N}\times\mathbb{N}\rightarrow\BExp(\Id)$. More precisely, $choose(ids,min,max)$ converts a set of \Id s into a Boolean expression, where at least \textit{min} and at most \textit{max} \Id s can be satisfied simultaneously.

For $x\in\Id(m)$ and if $x$ denotes an interface, we continue the definition of \rewrite as follows:
\begin{align}
\begin{split}
rewrite(x=0,m) &=\neg x \wedge \bigwedge_{i\in impls'(x,m)}{\neg i} \\
rewrite(x>0,m) &=x \wedge \bigvee_{i\in impls'(x,m)}{i} \\
rewrite(x=1,m) &=x \wedge \bigxor_{i\in impls'(x,m)}{i} \\
rewrite(x>=\const,m) &= x \wedge \textit{choose}(impls'(x,m),\const,\mid impls'(x,m) \mid) \\
rewrite(x>\const,m) &= x \wedge \textit{choose}(impls'(x,m),\const + 1,\mid impls'(x,m) \mid)
\end{split}
\end{align}

\paragraph{Semantic function}
The propositional semantics of a model $m\in\Cdl$ is just an intersection of the propositional denotations of all the nodes, similar to the full semantics. However, we (have to) leave out \texttt{legal\_values} as already pointed out. Furthermore, we need some more context (the current model $m\in\Cdl$) for the \textit{rewrite} function, since our propositional semantic domain is not capable of carrying enough information any more.

\begin{equation}
\begin{split}
\pCdlDenotation m = &
\left(\bigcap_{n\in m} \pNodeDenotation{ n, m } \right)
\cap
\left(\bigcap_{n\in m} \pFlavorDenotation{ n, m } \right)
\cap
\left(\bigcap_{n\in m} \pCalculatedDenotation{ n, m } \right)
\cap
\left(\bigcap_{n\in m} \pInterfaceDenotation{ n, m } \right)\\
& \cap
\left\{~ c\! \in\! \Confs_p \mid c_p(x)=0 \text{ for all } x\in \Id\setminus\Id(m)
~\right\}
\end{split}
\end{equation}

The semantics of a node is a set of all configurations that can satisfy its
constraints. For the propositional version, we introduce the macro $\CTC_p=\forall e\in ai\cup req.(e\in dom(rewrite) \rightarrow \eval(rewrite(e,m),c_p)))$. 

\begin{equation}
	\pNodeDenotation{(n,p,\_,ai,req,\_,\_,\_,\_),m} =  \{ c_p \in \Confs_p \mid c_p(n) \rightarrow c_p(p) \wedge CTC_p \}
\end{equation}

We continue with the denotation of nodes according to their flavor property:
\begin{align}
\begin{split}
\pFlavorDenotation{(n,p,(\FlNone|\FlData),\_,\_,\_,\_,\_,\_),m}&=\{c_p\in\Confs_p \mid c_p(p) \wedge CTC_p \rightarrow c_p(n)\} \\
\pFlavorDenotation{(n,p,(\FlBool|\FlBoolData),\_,\_,\_,\_,\_,\_),m}&=\Confs_p
\end{split}
\end{align}

Similarly, we define the denotation of the \calculated\ property. For $cl\neq\bot$:
\begin{align}
\begin{split}
\pCalculatedDenotation{(n,p,\_,\_,\_,cl,\_,\_,\_),m}=\{c_p\in\Confs\mid c_p(p)& \wedge CTC_p \rightarrow \\
&c_p(n)=\eval(\textit{rewrite}(cl,m),c_p)\}
\end{split}
\end{align}

Finally, the propositional denotation of interfaces can be declared as follows:
\begin{align}
\begin{split}
\pInterfaceDenotation{(n,\_,\_,\_,\_,\_,\_,\textit{interface},\_),m}=\{c_p\in\Confs_p \mid & c_p(p) \wedge CTC_p \rightarrow\\
 & c_p(n)=\eval(\bigvee_{i\in impls'(x,m)}{i},c_p) \}
\end{split}
\end{align}

\bibliographystyle{splncs03}
\bibliography{cdl_semantics}

\begin{thebibliography}{1}
\providecommand{\url}[1]{\texttt{#1}}
\providecommand{\urlprefix}{URL }

\bibitem{berger.ea:2012:tr}
Berger, T., She, S., Lotufo, R., Wasowski, A., Czarnecki, K.: Variability
  modeling in the systems software domain. Tech. Rep. GSDLAB-TR 2012-07-06,
  Generative Software Development Laboratory, University of Waterloo (2012),
  available at \url{http://gsd.uwaterloo.ca/tr/vm-2012-berger}

\bibitem{berger2010variability}
Berger, T., She, S., Lotufo, R., Wasowski, A., Czarnecki, K.: Variability
  modeling in the real: A perspective from the operating systems domain. In:
  International Conference on Automated Software Engineering (ASE) (2010)

\end{thebibliography}

\end{document}